\documentclass[pdflatex, iicol]{sn-jnl}
\usepackage{biblatex} %Imports biblatex package
\usepackage{graphicx} % Required for inserting images
\usepackage{algorithm}
\usepackage{booktabs}
\usepackage{enumerate}
\usepackage{amssymb}
\usepackage{mathtools}
\usepackage{comment}
\usepackage{hyperref}
\usepackage{xcolor}
\usepackage[noend]{algpseudocode}
\usepackage{svg}

\usepackage[english]{babel}
\usepackage{csquotes}

\usepackage{amsthm}
\theoremstyle{definition}
\newtheorem{definition}{Definition}[section]

\newcommand{\CC}{C\nolinebreak\hspace{-.05em}\raisebox{.4ex}{\tiny\bf +}\nolinebreak\hspace{-.10em}\raisebox{.4ex}{\tiny\bf +}}
\def\CC{{C\nolinebreak[4]\hspace{-.05em}\raisebox{.4ex}{\tiny\bf ++}}}

\begin{document}

\title[Article Title]{Exploring the non-uniqueness of node co-occurrence matrices of hypergraphs}

\author*[1,2]{\fnm{Timothy} \sur{LaRock}}\email{larock@princeton.edu}

\author[1,3]{\fnm{Renaud} \sur{Lambiotte}}

\affil[1]{\orgdiv{Mathematical Institute}, \orgname{University of Oxford}, \orgaddress{\city{Oxford}, \country{UK}}}

\affil[2]{\orgdiv{CIS Group, Department of Civil and Environmental Engineering}, \orgname{Princeton University}, \orgaddress{\city{Princeton}, \state{New Jersey}, \country{USA}}}

\affil[3]{\orgname{Turing Institute}, \orgaddress{\city{London}, \country{UK}}}

\abstract{Hypergraphs extend traditional networks by capturing multi-way or group interactions. Given the complexity of hypergraph data and the wide range of methodology available for pairwise network analysis, hypergraph data is often projected onto a weighted and undirected network.
The simplest of these projections, often referred to as a node co-occurrence matrix, is known to be non-unique, as distinct non-isomorphic hypergraphs can produce the same weighted adjacency matrix. This non-uniqueness raises important questions about the structural information lost during the projection and how to efficiently quantify the complexity of the original hypergraph.
Here we develop a search algorithm to identify all hypergraphs corresponding to a given projection, analyze its runtime, and explore its parallelisability. Applying this algorithm to projections derived from a random hypergraph model, we characterize conditions under which projections are non-unique. Our findings provide a new framework and set of computational tools to investigate projections of hypergraphs.}

\keywords{}

\maketitle

\section{Introduction}
Hypergraphs are a useful representation for interaction data featuring multi-way or group interactions. As a generalization of pairwise network representations, hypergraphs are more flexible as mathematical objects since they can represent interactions among groups of arbitrary sizes. In their most general form, including both parallel edges and repeated nodes, hypergraphs are multi-sets of multi-sets, and are often represented numerically by sparse high-dimensional tensors or incidence matrices \cite{ouvrard_adjacency_2018, aksoy_scalable_2024}. Throughout this paper, we will focus on \emph{simple hypergraphs}, meaning hypergraphs in which no hyperedges occur more than once (no parallel or multi edges) and no node occurs more than once in a given hyperedge (no self-loops). Simple hypergraphs can be represented formally as sets of sets. 

\begin{figure*}
    \centering
    \includegraphics[scale=0.24]{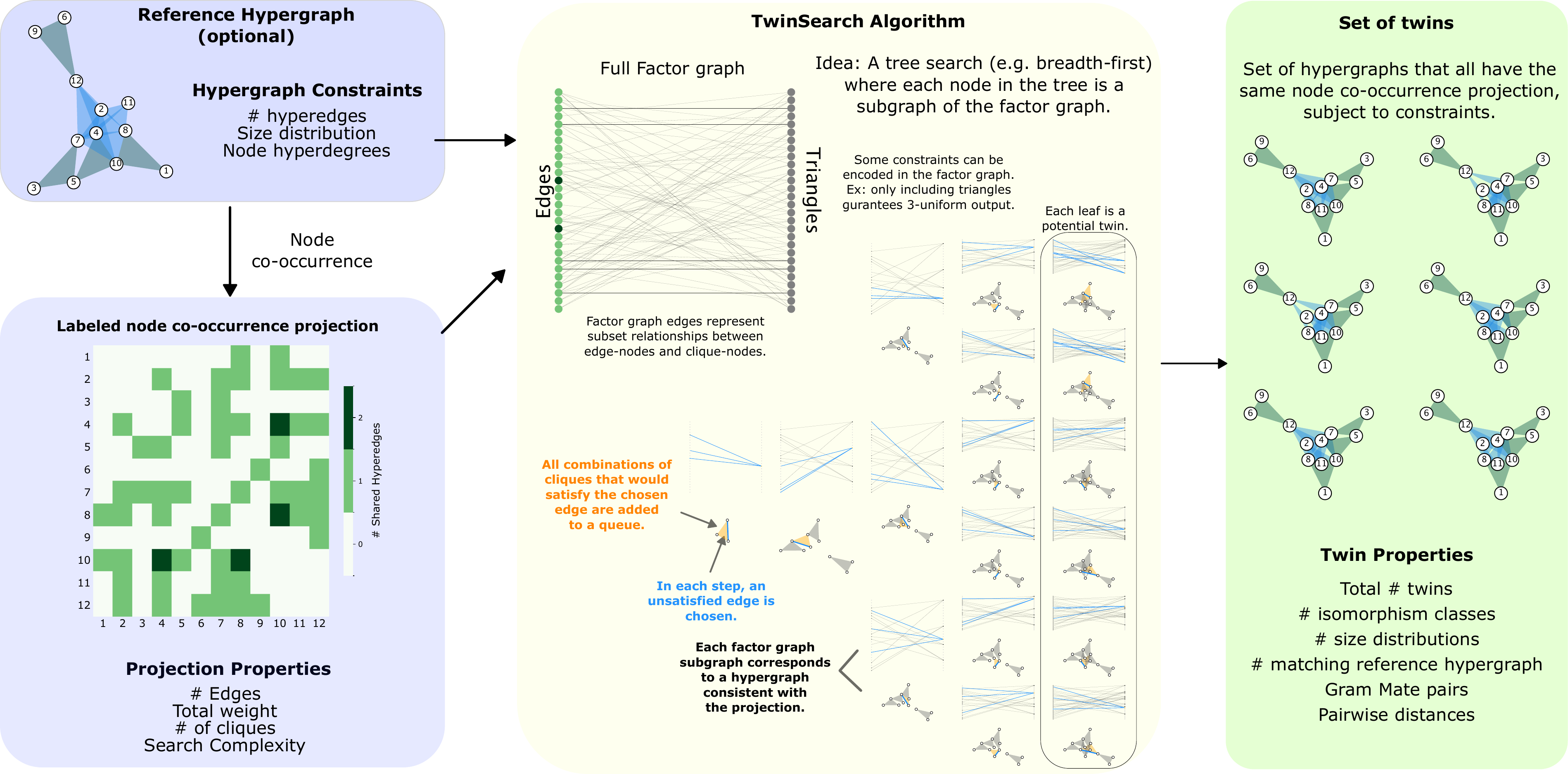}
    \caption{Overview of node co-occurrence projections of hypergraphs and our TwinSearch framework. The left panel shows a node co-occurrence projection matrix, optionally computed from a reference hypergraph. This forms the input to the TwinSearch procedure, shown in the middle panel and detailed in pseudocode in Algorithm~\ref{alg:dfs}. We first represent the projection as a factor graph, a bipartite graph computed from an adjacency matrix where one set of nodes represents pairwise edges and the other represents cliques \cite{young_hypergraph_2021}. In this example, we limit to only 3-cliques or triangles, which will in turn limit the output to $3$-uniform hypergraphs. The TwinSearch procedure iteratively searches subgraphs of the factor graph that are consistent with the input projection matrix and any user-imposed constraints, such as a desired hyperedge size or node hyperdegree distribution based on the reference hypergraph. The output of the search is a set of hypergraphs that all realize the input node co-occurrence projection that we call the \emph{set of twins}.}
    \label{fig:overview}
\end{figure*}

The representational complexity of hypergraphs, itself a consequence of their combinatorial complexity, sometimes leads researchers to simplify hypergraph data by projecting it back onto the set of nodes, creating an undirected dyadic multigraph where a link is placed between two nodes for each hyperedge they share \cite{battiston_networks_2020, carletti_random_2020, carletti_random_2021, ferraz_de_arruda_phase_2021, salova_analyzing_2021, gallo_higher-order_2024}. This projection technique can be applied to both bipartite (or two-mode) data and group interaction data, where it is sometimes referred to as the hypergraph adjacency matrix or weighted clique expansion \cite{everett_dual-projection_2013, kirkland_two-mode_2018, aksoy_hypernetwork_2020, aksoy_scalable_2024}.\footnote{We note that there is sometimes confusion in the literature, and that this nomenclature is not unique to this specific object. Moreover, there exist alternative ways to define the node projection of a hypergraph. See for example \cite{banerjee_spectrum_2021, carletti_random_2021, ferraz_de_arruda_phase_2021}.} This matrix is a convenient representation of hypergraph data because it corresponds to an adjacency matrix describing a weighted and undirected network, facilitating the direct application of pairwise network based methods to the analysis of group interaction data. Using the hypergraph adjacency matrix also reduces the largest dimension from the number of hyperedges (as in an incidence matrix) down to the number of nodes.

In this paper, we will refer to this matrix as the labeled node co-occurrence projection matrix of a hypergraph and define it as follows:

\begin{definition}[Labeled node co-occurrence projection]
    \label{def:projection}
    Given a hypergraph $H=(V, \mathcal{E})$ with $n=|V|$ nodes labeled $1,\dots,n$ and $m=|\mathcal{E}|$ hyperedges labeled $1,\dots,m$, the weighted node co-occurrence projection is a matrix $W_{u v}\in \mathbb{Z}^{n\times n}_{\geq0}$ with off-diagonal entries containing the number of hyperedges in which the pair of nodes $(u,v)$ co-occur:
    
    $$W_{u v} \coloneqq |\{(u,v) \subset e\}|_{ e \in \mathcal{E}}\textrm{.}$$

    The co-occurrence projection may or may not include diagonal entries, which correspond to node hyperdegrees in a hypergraph context and self-loops in a network context.
    If we represent the hypergraph $H$ as an $m \times n$ incidence matrix $A$, where $A_{\alpha u}=1$ indicates that the hyperedge $\alpha$ contains the node $u$, we can write the matrix defined above as:
    $$W \coloneqq A^TA\textrm{,}$$
    when self-loops are included, and
    $$W \coloneqq A^TA - D\textrm{,}$$
    when self-loops are excluded.
    Here $D$ is the $n\times n$ diagonal matrix containing node hyperdegrees (the number of hyperedges in which the node occurs). Throughout this work, we will consider co-occurrence projections both with and without hyperdegrees on the diagonal.\footnote{We note that throughout this paper we will use $W_{uv}$, $W_e$ (with index $e=(u,v)$ assumed), and simply $W$ interchangeably depending on context. Each refers to a node co-occurrence projection matrix.}
\end{definition}

It is known that a node co-occurrence projection is not in general unique to a single hypergraph \cite{kirkland_two-mode_2018, everett_dual-projection_2013, salova_analyzing_2021, ferraz_de_arruda_phase_2021}. In this paper we explore sets of hypergraphs that correspond to the same node co-occurrence projection. For example, Figure~\ref{fig:overview} shows a node co-occurrence matrix that can be realized by at least six different hypergraphs. For the sake of precision and ease of reference, we define sets of \emph{twin hypergraphs} as follows:

\begin{definition}[Set of twin hypergraphs]
    \label{def:twins}
    Given a node co-occurrence projection $W_{u v}$, the set of \emph{twin hypergraphs} associated to $W_{u v}$ is a set of hypergraphs $T\coloneqq \{h_1, \dots, h_{\ell
    } \}$ such that $W_{uv}$ is the node co-occurrence projection for each $h_i$ and for each distinct pair $i$ and $j$ we have $h_i \neq h_j$, meaning that the two labeled hypergraphs do not have the same incidence matrices.
\end{definition}

As noted in Definition~\ref{def:projection}, we can choose whether to consider the diagonal elements in $W_{uv}$ and it turns out that this choice can change the set of twins substantially. In the most general case, which we will refer to as the ``unrefined'' set of twins, we consider all hypergraphs to be in the set of twins if the off-diagonal entries of their projections match the off-diagonal entries of $W_{uv}$, effectively setting the diagonal to 0. In this unrefined set, properties such as the number of hyperedges, the hyperedge size distribution, and the node hyperdegrees are all unconstrained beyond the structure encoded in the off-diagonal entries of the projection matrix.

We will consider three refinements of the set of twins. First, we refine the set of twins to those hypergraphs from the unrefined set whose projections also have the same non-zero diagonal entries as the input projection, meaning the hypergraphs all have the same labeled node hyperdegree distribution. Beyond the diagonal, we will also refine the set of twins by constraining the number of hyperedges to be exactly $m$, or the hyperedge size distribution to be $k$-uniform, meaning all hyperedges have exactly $k$ nodes. For this to make sense, we need as input not only the node co-occurrence matrix of interest, but also a fixed number of hyperedges and/or hyperedge size distribution that can realize it, usually obtained via a reference hypergraph.

For any set of twins, we can also define the set of \emph{non-isomorphic twins} to be the same set, but with only one representative from each \emph{hypergraph isomorphism class}. We say two hypergraphs are isomorphic if there is a mapping between their node labels such that applying the mapping to one hypergraph makes it indistinguishable from the other. We can compute a partition of a set of twin hypergraphs into disjoint subsets based on isomorphism, then count how many subsets (also known as isomorphism or equivalence classes) we have in our partition. Hypergraphs can be tested for isomorphism by running graph isomorphism tests on their bipartite representations \cite{salova_analyzing_2021}.

Sets of twins are closely related to the concept of \emph{Gram Mates}, analyzed recently by Kirkland \cite{kirkland_two-mode_2018} and then Kim and Kirkland \cite{kim_gram_2023}. Gram Mates are pairs of binary rectangular matrices (including both incidence matrices of hypergraphs and bipartite adjacency matrices) with the same projection in both directions. Formally:
\begin{definition}[Gram Mates ]%\cite{kim_gram_2023}]
    \label{def:gram-mates}
    A pair of binary rectangular matrices $(A,B)\in \{0,1\}^{M \times N}$ are \emph{Gram Mates} if the following conditions hold:
    \begin{enumerate}[(i)]
        \item $AA^T = BB^T$,
        \item $A^TA = B^TB$, and
        \item $A\neq B$.
    \end{enumerate}
\end{definition}

Condition (ii) above is the node co-occurrence projection matrix with hyperdegrees on the diagonal (see Definition~\ref{def:projection}). Kirkland showed that as the number of nodes or the number of hyperedges goes to infinity, the asymptotic probability of randomly finding a pair of Gram Mates vanishes \cite{kirkland_two-mode_2018}. Later, Kim and Kirkland showed that Gram Mates can be constructed by changing the signs of their singular values and defined some infinite families of Gram Mates, including non-isomorphic Gram Mates \cite{kim_gram_2023}.

While Gram Mates are closely related to the motivations of this paper, the additional requirement that the line graph is also equivalent (condition (i) in Definition~\ref{def:gram-mates}) is stronger than necessary to pose a problem for network science methods. We note that the distinction between sets of twins and pairs of mates immediately suggests one plausible route for remediation, which is to study not only the node co-occurrence projection, but also the weighted line-graph projection, which is always equal in Gram Mates, but not necessarily equal among sets of twins. We also note that Aksoy et al. have found that tensor centrality methods can distinguish Gram Mates, suggesting that tensor methods may be a promising direction for future research \cite{aksoy_scalable_2024}.

The goal of our work is to understand the relationship between node co-occurrence projections and their sets of twin hypergraphs. We will take an algorithmic and numerical approach by describing and applying an algorithm, called \texttt{TwinSearch}, that accepts a node co-occurrence projection matrix as input and enumerates all hypergraphs that can realize the input projection, \emph{i.e.}, the set of twins. We then analyze these sets of twins first in terms of their overall size, then their diversity with respect to properties such as hyperedge size distributions, node hyperdegree distributions, pairs of Gram Mates, and pairwise distances. Figure~\ref{fig:overview} shows an overview of our approach to enumerating and analyzing sets of twins.

The motivation of our work is to provide a framework for researchers to assess whether, and to what extent, the non-uniqueness of node co-occurrence projection matrices may pose a challenge for projection-based methods in capturing the complex structure of hypergraphs. The main risk involved when using a node co-occurrence projection matrix is finding that structurally distinct hypergraphs, including non-isomorphic pairs with differing numbers of hyperedges, can produce identical projections, and thus cannot be distinguished by methods relying on the projection to encode the hypergraph structure. For example, recent research in dynamical systems on hypergraphs has shown that non-isomorphic hypergraphs with isomorphic node co-occurrence projection matrices can exhibit different dynamical behaviors from one another and from their projection \cite{salova_analyzing_2021, ferraz_de_arruda_phase_2021}. Further, some methods for analysis of hypergraphs may also be sensitive to these issues. Methods that project hypergraph data to node co-occurrence matrices may not be sensitive to the replacement of the original hypergraphs with non-isomorphic counterparts \cite{gallo_higher-order_2024}.

Also related to our research are the papers by Young et al. \cite{young_hypergraph_2021} and Wang and Kleinberg \cite{wang_graphs_2024}, both of which study reconstruction of hypergraphs from pairwise projection matrices. The essential difference between their work and ours is that we are interested not in inferring hyperedges from pairwise data, but in the distribution or ensemble of hypergraphs corresponding to the same labeled node co-occurrence projection matrix. Directly related to this latter point is the recent paper by Agostinelli et al. \cite{agostinelli_higher-order_2025} that sampled cliques of varying sizes to construct null models for understanding hypergraph dissimilarity measures. We will discuss this work in more detail in the Results.
\vspace{-0.2cm}
\subsection*{Contributions}
In the remainder of this paper, we describe our \texttt{TwinSearch} algorithm to enumerate all hypergraphs associated to a labeled node co-occurrence projection matrix. We then use this algorithm to study the extent to which non-uniqueness of node co-occurrence projection matrices emerges in a random setting and the relationship between basic properties of hypergraphs and the uniqueness of their node projections and \emph{vice versa}.

The paper is organized as follows. In the next section we give our \texttt{TwinSearch} algorithm for finding the set of twin hypergraphs associated to a node co-occurrence projection. We briefly analyze the runtime and describe how it can be parallelized. We then use the algorithm to characterize the parameter space of a simple random hypergraph model, studying the size of the set of twins and related quantities. Along the way, we also consider some basic combinatorial upper-bounds on the possible number of twins and discuss the implications of fixing different hypergraph properties, namely the hyperedge size and node hyperedgeree distributions. Then we analyze sets of twins through measures of pairwise hypergraph dissimilarity and finally conclude with a discussion of some practical implications of our results and open avenues for future research.
\vspace{-0.2cm}
\section{Enumerating sets of twins}
\label{sec:methods}
Let us assume that we are given as input a node co-occurrence projection matrix $W_{u v}$ where each off-diagonal entry contains the number of hyperedges in which the pair of distinct nodes $(u,v)$ co-occur (see Definition~\ref{def:projection}). We need not assume anything else to define the algorithm. However, it is worth reiterating that we will sometimes be particularly interested in the diagonal of the input projection, and in later sections we will also focus on \emph{$k$-uniform} hypergraphs, where all hyperedges have $k$ nodes, and will use non-uniform to refer to hypergraphs with hyperedges of arbitrary sizes.

We now describe and analyze our algorithm for enumerating sets of twins from node co-occurrence projection matrices, shown in pseudocode in Algorithm~\ref{alg:dfs}. We are given a weighted node co-occurrence projection matrix $W$. The output is the set of twins, meaning all hypergraphs that have labeled projections equal to $W$ (see Definition~\ref{def:twins}). The idea of the algorithm is to enumerate hypergraphs with equivalent projections by iteratively selecting an edge in $W$, searching over all of the possible combinations of cliques in $W$ that could satisfy the edge, then repeating until either a hypergraph satisfying $W$ is found or no further hyperedges can be added without violating a constraint. This defines a tree search over partial hypergraphs consistent with $W$, which can be implemented equivalently as either depth-first or breadth-first. Since every partial hypergraph in the search is guaranteed to be unique, the procedure can be parallelized (discussed below).

The most generic version of the algorithm will simply output all simple hypergraphs consistent with the input, but we may also wish to filter the set of twins based on some refinement criteria, \emph{e.g.}, representatives of twins that are not isomorphic to one another, or twins that also have equivalent line graph projections (Gram Mates, see Definition~\ref{def:gram-mates}). This filtering can occur during the execution of the algorithm, which may allow early termination of some branches at the expense of simple parallelization, or at the end using pairwise comparisons.

\begin{algorithm}
	\caption{Enumerate all hypergraphs on $n$ nodes that have the same labeled pairwise projection $W$.}
	\label{alg:dfs}
	\begin{algorithmic}[1]
    \Procedure{TwinSearch}{$W$}
        \State twins $\gets \emptyset$
        \State $F(V_e, V_c, \eta_e) \gets \texttt{FactorGraph}(W)$
        \State Choose edge $e \in V_e$ s.t. $W_{e} > 0$
        \Statex {\verb!  // Init. stack with combinations !}
        \Statex {\verb!  //  of cliques that would satisfy e !}
        \State $S\gets \emptyset$
        \For{$\delta \in {\binom{\eta_e }{W_{e}}}$}
            \State $S\gets S \cup \delta$
        \EndFor
	    \While {$S \neq \emptyset$ }
            \Statex {\verb!     // Get next partial hypergraph!}
            \State $T \gets \texttt{pop}(S)$
            \State Compute projection $W'$ of $T$
            \State $R \gets W - W'$
    		\If {$\sum R = 0$}
                \Statex {\verb!        // Twin found!}
                \State twins $\gets$ twins $\cup$ $T$
                \Statex {\verb!        // Check for line graph equiv.!}
                \Statex {\verb!        // and/or isomorphism!}
            \Else
                \Statex {\verb!        // Otherwise, continue search!}
                \State Compute $F'$ with $V'_c=V_c\setminus T$
                \State Choose an edge-node $e$ with $R_e > 0$
                \For {$\delta \in \binom{\eta'_e}{W'_e}$}
                    \State $\texttt{push}(S, T\cup \delta)$
                \EndFor
            \EndIf
		\EndWhile
    \Return twins
    \EndProcedure
	\end{algorithmic}
\end{algorithm}

We implement our algorithm using the \emph{factor graph} representation of an adjacency matrix as defined in \cite{young_hypergraph_2021}. The factor graph is a bipartite graph where one set of nodes corresponds to the edges from the adjacency matrix (edge-nodes), and the second set of nodes correspond to all of the cliques in the projection (clique-nodes). Importantly, the factor graph includes all cliques from the adjacency matrix, not only maximal cliques. In our application, the adjacency matrix of interest is the node co-occurrence projection $W$. Since the weights of $W$ do not play a role in the construction of the factor graph, we can consider its binarized equivalent. The factor graph is then defined as a tuple $F=(V_e, V_c, \eta_e)$ computed from $W$, where $V_e$ is the set of edge-nodes (non-zero entries $e=(u,v)$ in $W_{uv}$, $u < v$), $V_c$ is the set of clique-nodes, and $\eta_e$ is the adjacency list indexed by edge-nodes, where each index contains the list of clique-nodes of which the index edge-node is a subset. We show example factor graphs in Figure~\ref{fig:overview}.

Using the factor graph, we can view our search as a constraint satisfaction problem where we search for subgraphs of the factor graph satisfying $W$. That is, we want to find subgraphs of the factor graph such that each edge-node $e=(u,v)$ has $W_{uv}$ neighbors. We also use the factor graph to impose further constraints on the search. For example, if we are interested in $k$-uniform hypergraphs, we can include in the factor graph only cliques of size $k$ nodes, potentially reducing the search space substantially. Similarly, if we wish to find hypergraphs with a specific hyperedge size distribution $\mathcal{M}_k$, we can impose that the sum of degrees for clique-nodes of each size $k$ is exactly $\mathcal{M}_k\cdot k$ and reject any partial hypergraphs that violate this constraint. If $W$ has non-zero entries along the diagonal, we can also track these throughout the algorithm by including the individual nodes alongside the set of edge-nodes in the factor graph, or by simply maintaining the node hyperdegree distribution separately.

The algorithm proceeds as follows. First, we construct the factor graph $F=(V_e, V_c, \eta_e)$ from $W$. We then choose an edge-node $e$ with positive $W_{e}$.
\footnote{In practice, we choose this edge in the fastest way possible, thus making the order of execution arbitrary. One can impose an ordering on the satisfaction of edges to achieve a specific goal. For example, to minimize the size of the search tree, we can satisfy fully-determined edges first by always choosing the edge that minimizes $c(e) = \binom{|\eta_e|}{W_{uv}}$. When $c(e) = 1$, all of the clique-nodes attached to edge-node $e$ in the factor graph must be included in a hypergraph that satisfies the constraints of $W$. By satisfying edges with $c(e)=1$ immediately, we avoid adding a layer of depth later in the search tree when it may be wider, thus adding unnecessary nodes, and start from the smallest possible problem. However, in the worst case computing this constraint requires a loop over the pairwise edges at every node in the search tree, which may not always be practical.}
Every combination of $W_{e}$ clique-node-neighbors from $\eta_e$, each corresponding to a partial hypergraph that satisfies $e$, is added to a stack (or queue) before entering the main loop.

In the main loop, we pop a partial hypergraph off of the stack, compute its projection $W'$ and the remaining edges $R=W-W'$. If the sum of $R$ is $0$, then the current hypergraph has satisfied all of the edge-nodes, meaning it is consistent with the input. At this point we may choose to either immediately check this hypergraph against other hypergraphs we have discovered thus far for filtering, or we may simply add it to the set of all hypergraphs corresponding to this projection to be checked later.

If the sum of $R$ is non-zero, we choose another edge-node $e$ with positive $W'_e$ and repeat. If at any point we find that the constraints cannot be satisfied (for example, we have edges with $R_e > 0$, but $|\eta_e|=0$) we add nothing to the stack and move on, ending the search along the current branch.

Once the algorithm returns, we are left with a set of hypergraphs on $n$ nodes that all have node co-occurrence projection matrices equal to the input matrix $W$, subject to our imposed constraints. If necessary, the last step is to compare all of the hypergraphs to one another and keep only one representative for each isomorphism class by running the graph isomorphism test on the bipartite representations (following the same procedure as \cite{salova_analyzing_2021}). We may also check for any pairs of Gram Mates by comparing the line graph representations of each pair of hypergraphs. If the input $W$ does not correspond to any simple hypergraph (subject to constraints), the algorithm will simply finish with the set of twins still empty.
\vspace{-0.2cm}
\subsection{Parallelization}
The \texttt{TwinSearch} algorithm searches a tree where each node corresponds to a partial hypergraph (see the middle panel of Figure~\ref{fig:overview} for a visual explanation). Each partial hypergraph is considered exactly once, thus each descendant is independent from all others. For this reason, the algorithm can be parallelized using any paradigm that allows for work to be added dynamically as the process unfolds. For example, in our implementation (available at \cite{larock_TwinSearch_2023}) we use a queue that can receive new partial hypergraphs as they are generated by processing previous nodes. At any point in the program this queue in effect represents the ``boundary'' of the search, meaning that nodes on the queue correspond to the deepest point reached on a specific branch of the search tree.
\vspace{-0.2cm}
\subsection{Computational Complexity}
\label{sec:complexity}
Since the algorithm is a tree search, to compute a bound on the computational complexity we can first compute the complexity of the initialization, then the complexity of processing an individual node, then finally bound the total number of nodes in the search tree.

In the initialization phase we compute the factor graph representation, which involves enumerating all of the cliques up to the maximum size $k$. We will use $\mathcal{C}_i$ to denote the number of cliques of size $i \in 2,\dots,k$. We know that $\mathcal{C}_2$ is just the set of edges. To get the 3-cliques, we can loop over each of the $\mathcal{C}_2$ edges and check for common neighbors with any of the other $n-2$ nodes. We can achieve this by summing the rows of the binarized projection matrix corresponding to each member of the clique; any entry in the resulting vector with value exactly $i$ is a common neighbor. This requires an $O(n\cdot i)$ vector addition for every clique. Thus to get the cliques up to size $k$, we can repeat this operation for increasing clique sizes for a complexity of $$\sum_{i\in 3\cdots k} \mathcal{C}_i\cdot n \cdot i\textrm{.}$$

Once the factor graph has been constructed, we then choose an edge $e$ in constant time and add $\binom{\eta_e}{W_{e}}$ nodes to the stack or queue, corresponding to the possible combinations of cliques connected to $e$ that are compatible with $W$. This is the main work that needs to be done at every node of the search tree.

We now turn to the size of the search tree. A trivial upper bound on the depth of the tree is $\mathcal{O}{(|E|)}$, meaning there is one layer for every edge in the projection. In practice, especially for sparse projections, we expect the depth to be smaller, since some edges will be satisfied incidentally through the satisfaction of earlier edges.

The maximum number of items that can be added to the stack is an upper bound on the width of each layer. This number is bounded by the maximum number of sets of cliques that could satisfy any edge $e$, computed on the full factor graph: $$C=\max\limits_{e}{\binom{|\eta_e|}{W_{e}}}.$$  This gives a trivial worst-case search tree size of $\mathcal{O}{(C^{|E|})}$, the case where every layer of the tree has the same width. A tighter bound on the worst case search tree size is simply the product of the widths at every edge:

\begin{equation}
\mathcal{O}(\prod\limits_{e}{\binom{|\eta_e|}{W_{e}}}).
\end{equation}\label{eq:worst-case}

Later we will show numerical results regarding the relationship between this worst-case running time and properties of the input, as well as wall-clock running time in our {\CC} implementation (see Figure~\ref{fig:clique-approximation}) \cite{larock_TwinSearch_2023}.

\section{Results}
In this section we give some results regarding the relationship between properties of a node co-occurrence matrix and its set of twins. In the first two subsections we focus on some basic analytical results to help contextualize the search for twins. We first give a combinatorial upper bound on the number of hypergraphs corresponding to a given projection and explain how this upper bound overcounts by ignoring the detailed topology of the projection matrix. Then, we discuss in detail the implications of considering the diagonal entries of the projection in the $k$-uniform and non-uniform cases.

In the remaining subsections we focus on numerical results. Our results center around computing sets of twins for projections that correspond to hypergraphs sampled from a modified $G_k(n,m)$ or Erd\H{o}s-R\'{e}nyi model for $k$-uniform hypergraphs, described below. We analyze how the size of the set of twins changes with the parameters of the model and search constraints, including both $k$-uniform and non-uniform hyperedge size distributions, and we discuss the concentration of pairs of Gram Mates among the projections. Then, we analyze the distribution of pairwise distance measures among sets of twin hypergraphs. Finally, we present results from sampling projections for a wider variety of parameters.

\subsection{Combinatorial Upper Bounds}
\label{sec:upper-bound}
We can begin to relate node co-occurrence projections to their sets of twin hypergraphs through combinatorial arguments. These arguments are closely related to our previous description of the size of the tree searched by the algorithm. We again assume we are given a node co-occurrence projection matrix $W$ on $n$ nodes. We further assume we are given a reference hyperedge size distribution, denoted $\mathcal{M}_k$.

First let us assume that the hypergraph is $k$-uniform, meaning $\mathcal{M}_k$ is non-zero only for $k$. We can immediately give an upper bound on the number of possible $k$-uniform hypergraphs corresponding to $W$ as:

$$\binom{\binom{n}{k}}{\mathcal{M}_k}.$$
The inner binomial coefficient corresponds to the number of possible $k$-hyperedges on $n$ nodes and the outer binomial counts the total number of hypergraphs on $\mathcal{M}_k$ $k$-uniform hyperedges. If we further know $\mathcal{C}_k$, the distribution of the number of cliques of each size $k$ in $W$ introduced previously, we can refine this to:

$$\binom{\mathcal{C}_k}{\mathcal{M}_k}.$$

We know that $\mathcal{C}_k \leq \binom{n}{k}$, therefore this second bound is always better except in the case where every possible clique on $k$ nodes exists in the projection.\footnote{We note that in this analysis the cliques are binary while the projection is weighted. Thus having all possible cliques in the projection does not necessarily fix the set of twins.}

\begin{figure}
    \centering
    \includegraphics[scale=0.55]{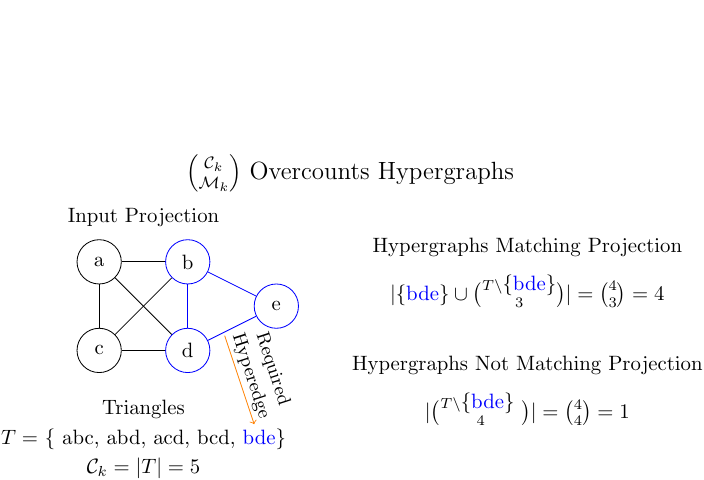}
    \vspace{0.1cm}
    \caption{Example describing how binomial coefficients overcount the number of hypergraphs associated to a projection in the 3-uniform case. Given the input projection on the left, when $\mathcal{M}_k=4$ hyperedges of size $k=3$, there is 1 extra hypergraph counted by the binomial coefficient $\binom{\mathcal{C}_k}{\mathcal{M}_k}$ that does not have the input projection as its node co-occurrence projection matrix.}
    \label{fig:overcounting}
\end{figure}

We expect that this quantity will overcount the hypergraphs corresponding to a specific projection, since we are using only the distribution of clique and hyperedge sizes, thus losing all information about the detailed structural relationships between pairs of nodes contained in the projection. To illustrate the potential for overcounting precisely, consider the example shown in Figure~\ref{fig:overcounting}, where we have a projection on $n=5$ nodes with $\mathcal{C}_3=5$ cliques on 3 nodes. To understand the overcounting problem, it is sufficient to consider the 3-uniform case with $\mathcal{M}_3=4$ hyperedges on 3 nodes. As illustrated in Figure~\ref{fig:overcounting}, any 3-uniform hypergraph corresponding to this projection must contain the triangle $\{\textrm{bde}\}$, otherwise the edges $\{\textrm{be}\}$ and $\{\textrm{de}\}$ will not appear. Since the remaining structure is a 4-clique, we have 4 ways of choosing 3 more triangles as hyperedges, each of which will correctly match the projection. However, the quantity $\binom{\mathcal{C}_k}{\mathcal{M}_k}$ does not disallow the hypergraph made up of the 4 triangles in the 4-clique, even though this hypergraph does not match the projection. Thus the binomial counts 1 extra hypergraph in this case and can overcount hypergraphs in general without further constraints.

If $\mathcal{M}_k$ instead corresponds to a non-uniform hypergraph, meaning it has non-zero entries for multiple values of $k$, we can also compute this upper bound by taking a product over the relevant values of $k$, using the distribution of clique sizes in $W$ as before:

\begin{equation}
\prod_{k \textrm{ s.t. } \mathcal{M}_k>0} \binom{\mathcal{C}_k}{\mathcal{M}_k}.
\end{equation}\label{eq:approximation}

We will refer to the product of binomials in Equation~\ref{eq:approximation} as the \emph{clique approximation} to the size of the set of twins. Each term in this product is susceptible not only to the overcounting described above, but also to overcounting by not respecting the weights in $W$ in other ways. For example, if we assume there is an edge in $W$ with weight 1 that participates in at least 1 triangle, the above product counts hypergraphs where that edge would appear both as a 2-hyperedge and in a 3-hyperedge, which would result in a weight larger than 1.

The question remains how many of those possible hypergraphs actually correspond to the input projection $W$? More complex counting arguments may be fruitful, however we elect to approach the problem numerically, using the \texttt{TwinSearch} algorithm to investigate the relationship between the weighted edge structure of a node co-occurrence projection and its corresponding set of twins. Later, in Figure~\ref{fig:clique-approximation}, we will show the relationship between the true number of twins with a given hyperedge size distribution $\mathcal{M}_k$ corresponding to a projection with a given clique size distribution $\mathcal{C}_k$ and the approximation given in Equation~\ref{eq:approximation}.

\begin{figure*}[!ht]
    \centering
    \includegraphics[scale=0.31]{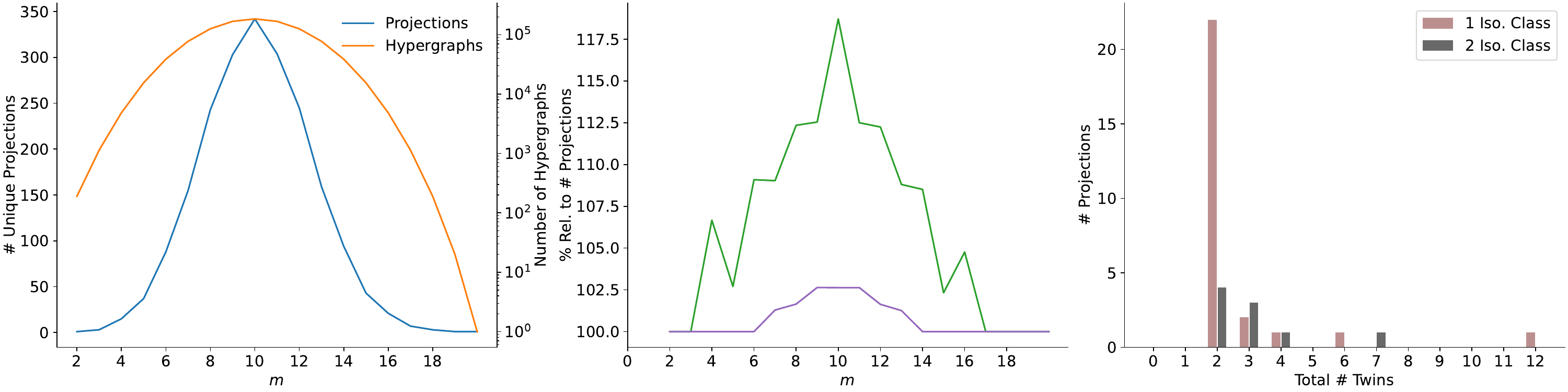}
    \caption{Left: Number of unique (up to isomorphism) node co-occurrence projections (blue, left-hand vertical axis) and $3$-uniform hypergraphs (orange, right-hand vertical axis) in $G_3(6,m)$ for increasing $m$. Middle: Percentage of $3$-uniform twins (green line) and isomorphism classes (purple line) corresponding to projections of hypergraphs with $m$ hyperedges. Right: Histogram counting the number of projections with a given number of $3$-uniform twins, broken down by whether projections have a total of 1 isomorphism class (brown bars) in the set of twins versus 2 (dark gray bars), the only values observed.}
    \label{fig:exhaustive-3-unif}
\end{figure*}

\vspace{-0.2cm}
\subsection{Dealing With the Diagonal}

The diagonal entries of the node co-occurrence projection correspond to node hyperdegrees. These hyperdegrees have different importance with respect to sets of twin hypergraphs depending on whether we are interested only in $k$-uniform hypergraphs or in all possible non-uniform hypergraphs. In particular, for $k$-uniform hypergraphs, the diagonal entries are redundant and fixed by the off-diagonal entries, meaning they are always equal among sets of twins. Formally, given a $k$-uniform hypergraph as an incidence matrix $A$ and its node co-occurrence projection $P=A^TA$, we have diagonal entries

$$P_{u,u} = \frac{\sum_{v \neq u} P_{u,v}}{k-1}\textrm{.}$$

This means that for any $3$-uniform incidence matrix $B$ that is a twin of $A$, we have 
$$ A^TA = B^TB \iff A^TA - D_A = B^TB - D_B\textrm{,}$$
where $D_A$ and $D_B$ are diagonal matrices that contain node hyperdegrees. Therefore, when we think about twins of $k$-uniform hypergraphs, we need not consider the diagonal entries.

However, this is not the case when we allow for non-uniform hypergraphs, since different size distributions among twins will result in different node hyperdegree distributions, meaning that we need to take some extra care in how we analyze sets of twins in non-uniform hypergraphs. The most trivial example is the individual 3-hyperedge and the triangle on 3 pairwise edges. In both cases the off-diagonal entries of the projection are exactly the same, but the hyperdegrees in the single 3-hyperedge case are 1, while in the pairwise edge case they are 2. However, the problem is more complex, as it can also occur when the same number of hyperedges are used. For example, the hypergraphs with edge sets $\{AD, CD, ABC\}$ and $\{AB, BC, ACD\}$ have the same off-diagonal projection, but nodes $B$ and $D$ have different hyperdegrees, since they replace one another in the isomorphism between the two hypergraphs.

There are various ways we can deal with this practically. For the results of this paper, we usually compute all of the twins, regardless of the diagonal, then present results broken down by whether the hyperdegree distributions match. As noted previously, we could also modify the algorithm to impose the constraint that the diagonal entries must match the input projection, first by stopping the search of a partial hypergraph if any diagonal becomes too large, then by rejecting any hypergraphs where a diagonal entry is too small.

\subsection{Case Study: Exhaustive Search}

We now turn to numerically analyzing sets of twins. We will focus our analysis on a modified Erd\H{o}s-R\'{e}nyi model for hypergraphs in which we choose $m$ hyperedges uniformly without replacement from $\binom{n}{k}$ possible hyperedges, rejecting hypergraphs where some node $u$ is not included in any of the $m$ sampled hyperedges (\textit{i.e.}, $u$ is a singleton), guaranteeing that the minimum hyperdegree of any node is 1. We choose to impose this constraint in this context because for our algorithm there is no difference between a hypergraph on $n$ nodes that includes $n_s$ singletons and a hypergraph on $n-n_s$ nodes, since all entries in the projection matrix involving any of the $n_s$ singleton nodes will be 0. 

In this section we will focus in detail on projections of hypergraphs from the $3$-uniform ensemble on $6$ nodes, denoted $G_3(6,m)$. We focus on these parameters because it is practical to enumerate all simple hypergraphs and their node co-occurrence projections across all values of $m$ between 2 and $\binom{6}{3} = 20$.

For each enumerated hypergraph, we compute its node co-occurrence projection, then run a pairwise isomorphism test between all projections and keep only one representative of each isomorphism class. The number of unique node co-occurrence matrices up to isomorphism for simple hypergraphs from $G_3(6,m)$ for each $m$ is shown in the left-hand plot of Figure~\ref{fig:exhaustive-3-unif} (blue line, left-hand vertical axis), as well as the total number of hypergraphs (ignoring our minimum-degree constraint) given by the binomial coefficient (orange line, right-hand vertical axis, log scale). We then run \texttt{TwinSearch} on each projection to enumerate the set of twin hypergraphs and study the properties of sets of twins as $m$ increases.

In Figure~\ref{fig:exhaustive-3-unif} we focus on $3$-uniform hypergraphs found by the algorithm. The middle plot shows the number of $3$-uniform twins as a percentage of the number of unique projections for each $m$ (shown in the left-hand plot). If every projection was realized by exactly one unique hypergraph, both lines would be flat at 100\%, as we see outside of the range from $m=4 \dots 17$ for all twins (green) and outside the range $m=7\dots 13$ for isomorphism classes (purple). The total number of twins reaches a peak of 15\% larger than the number of projections at $m=10$, while the total isomorphism classes is maximum for $m=9,10,11$ at around 2.5\% of the number of projections.

\begin{figure}
    \centering
    \includegraphics[scale=0.295]{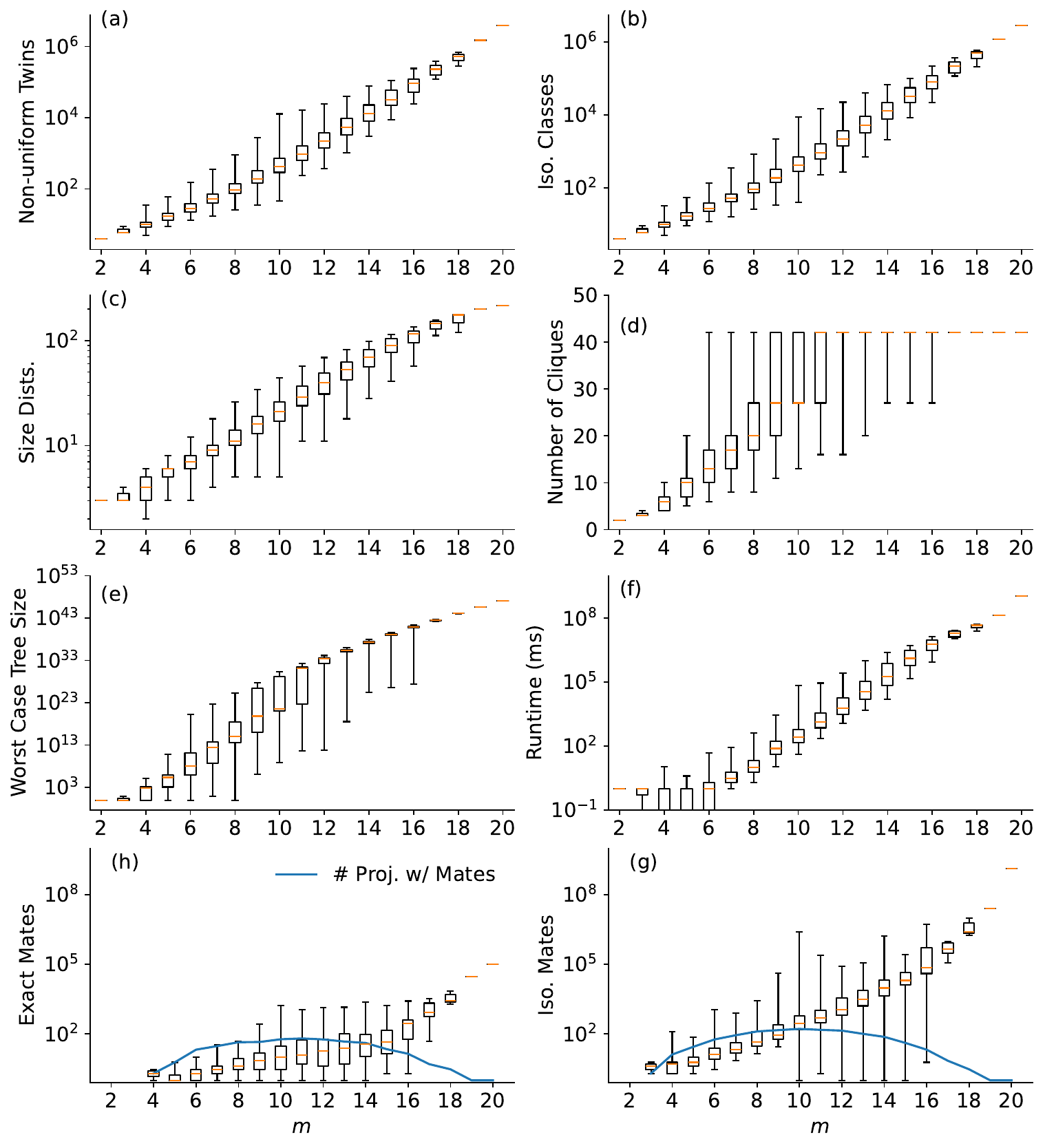}
    \caption{Relationships between number of hyperedges $m$ and the distribution of various quantities of interest, including: (a) the number of non-uniform twins; (b) the number of isomorphism classes among the set of twins; (c) the number of unique hyperedge size distributions; (d) the number of cliques in the projection; (e) worst case search tree size; (f) elapsed running time (note: vertical axis is cut off in log scale due to 0s corresponding to sub-millisecond runtimes); and pairs of exact (g) and isomorphic (h) Gram Mates. Each boxplot corresponds to the distribution of the quantity per projection from hypergraphs with $m$ hyperedges. Boxes span from 1st to 3rd quartiles, orange lines show median values, and whiskers represent minimum and maximum values. The blue lines in (g-h) show the number of projections with at least one pair of Gram Mates.}
    \label{fig:non-uniform-stats}
    \vspace{-0.35cm}
\end{figure}

\begin{figure*}[!ht]
    \centering
    \includegraphics[scale=0.25]{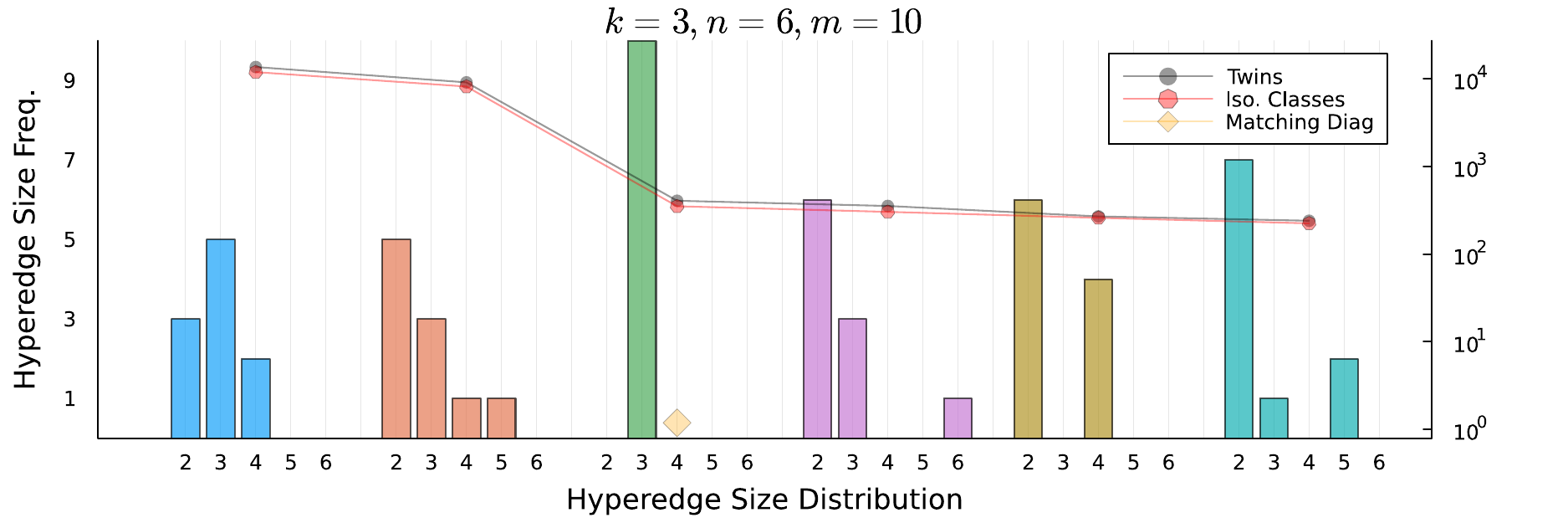}%
    \includegraphics[scale=0.25]{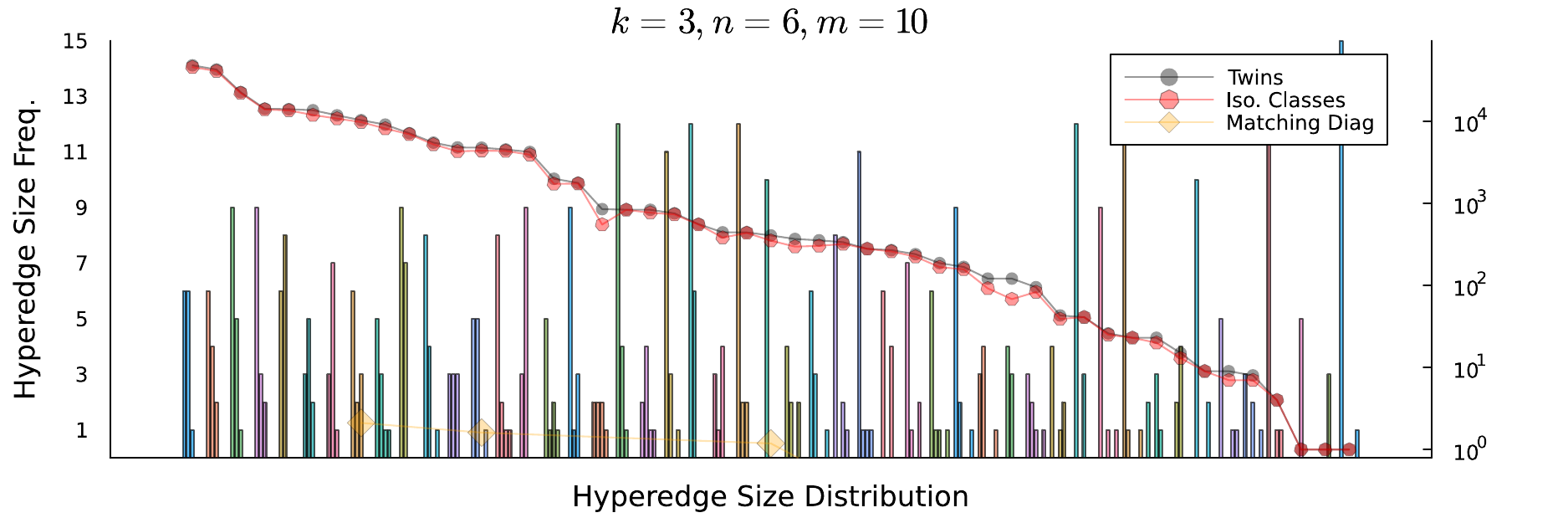}
    \caption{Hyperedge size distributions and twin set size distributions for non-uniform hypergraphs corresponding to the unique (up to isomorphism) node co-occurrence projections of hypergraphs from $G_3(6, 10)$. The left plot shows only statistics for hypergraphs in the twin set that have exactly $m=10$ hyperedges (matching the sampling procedure), while the right plot shows statistics over the whole set of twins. Each set of bars along the horizontal axis shows a hyperedge size distribution (indexed to the left-hand vertical axis). Each point in the line graphs is a statistic relating to that hyperedge size distribution: the gray line shows the total number of twins with that distribution, the red line shows the total number of isomorphism classes among those twins, and the yellow diamonds count the number of twin hypergraphs with hyperdegree distributions (diagonal entries) matching the input projection.}
    \label{fig:non-uniform-breakdown}
\end{figure*}

In the right hand plot of Figure~\ref{fig:exhaustive-3-unif}, we show histograms of the number of projections that have a given number of twins, broken down by the number of isomorphism classes among the twins. We show results for $m=10$, where all of the quantities in the left and middle plots are maximum, and we only show the distribution for projections that have more than 1 twin (since having only 1 twin implies only 1 isomorphism class). Our first observation is that while projections with exactly 2 twins are most likely to have only 1 isomorphism class, those projections also have the largest absolute number of projections with 2 isomorphism classes. Among the projections with 3 hypergraphs in their set of twins, more projections correspond to 2 different isomorphism classes than 1. Some projection matrices correspond to many hypergraphs that are all isomorphic to one another, \emph{e.g.}, the far right brown bar corresponds to a projection matrix with 12 hypergraphs in its set of twins, with each pair of hypergraphs isomorphic (hence 1 isomorphism class). In contrast, the gray bar furthest to the right shows that there is 1 projection that has 7 hypergraphs in its set of twins, with 2 isomorphism classes among them.

From these results analyzing only $3$-uniform sets of twins, we already see that the node co-occurrence projection operation can obscure substantial higher-order information and that this information loss is not homogeneous, even among projections sampled via the same random hypergraph model with the same parameters.

\subsection{Non-uniform Twins}
We now expand our analysis to sets of twins including non-uniform hypergraphs, where the hyperedge sizes are constrained only by the clique size distribution in the projection (equiv. the clique-nodes in the factor graph), \emph{i.e.}, hyperedges can in principle range in size from $2,\dots,n$. Figure~\ref{fig:non-uniform-stats} shows relationships between the number of hyperedges $m$ and distributions of various statistics among the full sets of twins associated to unique up to isomorphism node co-occurrence projections computed from hypergraphs in $G_3(6, m)$. The distributions are shown as box and whisker plots, where the limits of the box show the 1st and 3rd quartiles, the orange lines show the median, and the whiskers extend to the minimum and maximum values in the data.

The top row shows the distributions of the number of non-uniform twins (Figure~\ref{fig:non-uniform-stats}(a)) and isomorphism classes (b). The number of twins grows monotonically with $m$ and the distributions are not noticeably different, implying that a substantial proportion of twins are unique up to isomorphism.

Figure~\ref{fig:non-uniform-stats}(c) shows distributions of the number of hyperedge size distributions as $m$ increases. Again we observe fast growth up to more than 100 hyperedge size distributions for the twins corresponding to the projection at $m=20$. In Figure~\ref{fig:non-uniform-breakdown}, we zoom in on the distributions for $m=10$, showing a summary of the non-uniform hypergraphs corresponding to the unique projections from $G_3(6, 10)$. Along the horizontal axis we show hyperedge size distributions as bar charts indexed to the left-hand vertical axis. The plot on the left of Figure~\ref{fig:non-uniform-breakdown} is limited to only those twin hypergraphs that have exactly $m=10$ hyperedges, matching the input, while the right-hand plot does not limit the number of hyperedges. In both cases it is clear that there are many more possible hypergraph configurations corresponding to the node co-occurrence projections beyond the $3$-uniform hypergraphs we used as input.

\begin{figure*}[!ht]
    \centering
    \includegraphics[scale=0.28]{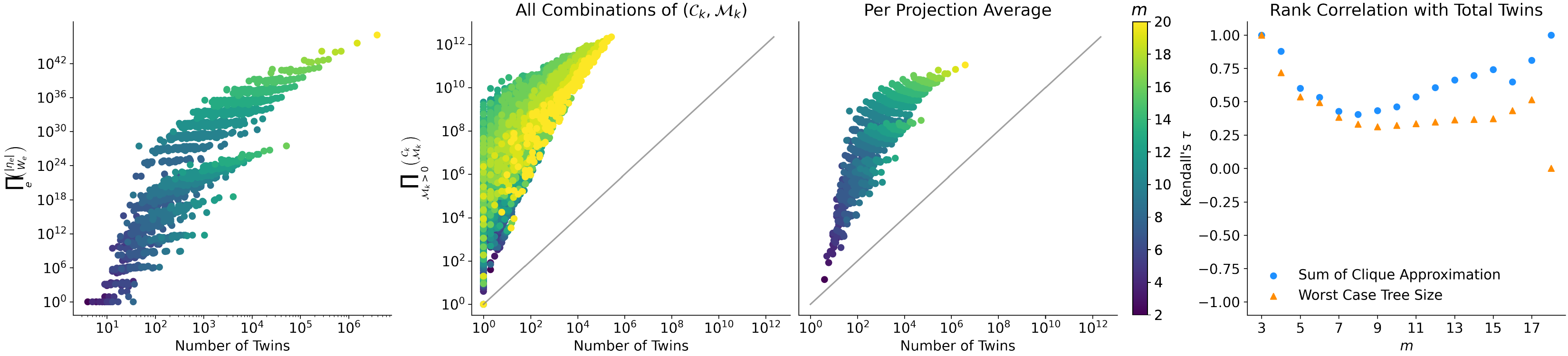}
    \caption{Left: Number of twins against worst case search tree size (see Eq.~\ref{eq:worst-case}) for unique up to isomorphism projection matrices from $G_3(6,m)$ (color in the first 3 plots corresponds to $m$). Middle: Numerical evidence of overcounting using the clique approximation (see Eq.~\ref{eq:approximation}, Fig.~\ref{fig:overcounting}, and surrounding discussion). The second plot shows results for every observed combination of clique size distribution $\mathcal{C}_k$ and hyperedge size distribution $\mathcal{M}_k$, while the third plot shows the average over all combinations for each projection. Right: Kendall's $\tau$ of rankings of projections by worst case search tree size (Eq.~\ref{eq:worst-case}, orange triangles) and the sum of the clique approximations for every combination of clique and hyperedge size distribution (Eq.~\ref{eq:approximation}, blue circles), both compared against the ranking of projections by their total number of non-uniform twins.}
    \label{fig:clique-approximation}
\end{figure*}

The line graphs in Figure~\ref{fig:non-uniform-breakdown} correspond to the right-hand vertical axis and show the number of twins (gray) and number of isomorphism classes (red) among twin hypergraphs corresponding to each hyperedge size distribution, per projection. The yellow line with diamond markers shows the number of hypergraphs with the corresponding hyperedge size distribution that have the same diagonal as the $3$-uniform input projection.

Focusing first on the set of twins with $m=10$ hyperedges (left-hand plot), we observe that 3-uniform twins are ranked third out of the 6 possible size distributions in terms of the total number of twins. We also note that in this case there is always exactly one 3-uniform twin with a matching diagonal. However, when we allow any number of hyperedges (right-hand plot of Figure~\ref{fig:non-uniform-breakdown}), we see a marked increase in the number of hyperedge size distributions. We also find that, contrary to the $m=10$ case, there are multiple distributions where the hyperdegrees on the diagonal match the input. Further, we note that there are some size distributions with many isomorphic twins, as well as some size distributions with exactly one twin.

Returning to Figure~\ref{fig:non-uniform-stats}, plot (d) shows the distribution of the total cliques of sizes $3,\dots,6$ in the input projections. With $m=6$, there are already some projections that have the maximum number of cliques, which we can compute as:
$$\sum_{k \in 3,\dots,6} \binom{6}{k} = 42\textrm{,}$$
and after $m=16$ all possible cliques are included in every projection, highlighting the importance of the weighted structure of the matrices.

Figure~\ref{fig:non-uniform-stats}(e) shows the distribution of the worst case search tree size (Equation~\ref{eq:worst-case}). As the number of hyperedges grows, the worst case complexity increases exponentially. Figure~\ref{fig:non-uniform-stats}(f) shows the elapsed runtime in our {\CC} implementation \cite{larock_TwinSearch_2023}. The runtime grows more slowly than the worst case search size, but still at an exponential rate with respect to $m$.

Figures~\ref{fig:non-uniform-stats}(g) and (h) show how the number of pairs of Gram Mates grows with the number of hyperedges. We identify pairs of Gram Mates in two ways. First, we check for \emph{exact} pairs, meaning that the labeled incidence matrices are compared against the conditions in Definition~\ref{def:gram-mates}. Each pair that meets those conditions is counted as an exact mate pair, shown in Figure~\ref{fig:non-uniform-stats}(g). However, we note that the labeling of hyperedges in twins computed by \texttt{TwinSearch} is an implementation detail. Therefore, we also count pairs of hypergraphs that we call \emph{isomorphic} Gram Mates, where the equalities in Definition~\ref{def:gram-mates} are changed to isomorphisms (Figure~\ref{fig:non-uniform-stats}(h)). For a given projection, the pairs of exact mates are a subset of the isomorphic mate pairs. We see from these distributions that the number of Gram Mate pairs increases steadily with $m$ and that the number of isomorphic pairs of Gram Mates is orders of magnitude larger than the number of exact Gram Mates.

Taken together, the results in Figures~\ref{fig:non-uniform-stats} and \ref{fig:non-uniform-breakdown} exemplify the diversity within the set of non-uniform twins corresponding to the node co-occurrence projection in terms of the number of hyperedges, the distribution of hyperedge sizes, node hyperdegrees, and Gram Mates. Our analysis so far shows that many key characteristics correlate with the density of hyperedges $m$. Notably, the worst case search tree size appears to correlate strongly with both the value of $m$ and, comparing across Figure~\ref{fig:non-uniform-stats}(a) and (e), the total number of non-uniform twins. This observation leads to our final analysis in this section, where we ask whether we can use approximations that do not require enumerating sets of twins to predict the number of twins associated to an individual projection.

In the leftmost plot of Figure~\ref{fig:clique-approximation}, we show the worst case search tree size (Equation~\ref{eq:worst-case}) against the total number of twins for each projection. We see a clear correlation between the two variables, meaning that a larger set of twins implies a larger search tree. This is on the one hand intuitive, since each twin is a leaf in the search tree, thus the number of twins is a trivial lower bound on the tree size. However, as can be seen from the order of magnitude of the worst case search tree size, there are large numbers of search tree nodes, both non-twin leaves and intermediate nodes, whose quantity could in principle have been less directly correlated to the true twin set size. What is powerful about the relationship between the number of twins and the search tree size is that it allows for the possibility of comparing projection matrices using just Equation~\ref{eq:worst-case}, which can be computed based only on the node co-occurrence matrix and its factor graph, without enumerating the full set of twins with \texttt{TwinSearch}.

We have also described another approximation for the number of twins, the clique approximation introduced in Section~\ref{sec:upper-bound}. In the middle two plots of Figure~\ref{fig:clique-approximation}, we show the relationship between the number of twins and Equation~\ref{eq:approximation}. The second plot shows results for all combinations of clique size distribution $\mathcal{C}_k$ and hyperedge size distribution $\mathcal{M}_k$ found in the data, while the third plot shows the average computed over all observed combinations of the two distributions per projection. Broadly, the approximation overcounts by orders of magnitude, but similarly to the worst case search tree size, there is an apparent correlation between the true number of twins and the approximation.

We now have evidence that our two approximations correlate with the true number of twins. However, the extent to which this information is practically useful remains to be seen. As a step in this direction, we use a simple rank correlation approach to understand whether these two approximations might be used to compare node co-occurrence projection matrices in terms of their number of twins, without running \texttt{TwinSearch}.

In the rightmost plot of Figure~\ref{fig:clique-approximation}, we show Kendall's $\tau$ rank correlations based on rankings of projection matrices. For each input hyperedge size $m$ such that there are at least 2 unique up to isomorphism node co-occurrence projections, we compute three rankings. First, we rank the projections by their total number of non-uniform twins; second, we rank by the the worst case search tree size (Eq.~\ref{eq:worst-case}); third, by the clique approximation (Eq.~\ref{eq:approximation}), summed over all observed combinations of $\mathcal{C}_k$ and $\mathcal{M}_k$. We then compute Kendall's $\tau$ rank correlation coefficients between the number of twins ranking and the two other rankings based on the approximations.

The rank correlation statistics for both approximations are generally positive and significant (below $0.05$), with some exceptions at the lowest and highest values of $m$, where there are relatively few projections (numerical coefficients and p-values are available in Appendix~\ref{app:kt-stats}, Table~\ref{app:tab:kt-stats}). While neither ranking is fully concordant with the number of twins, which would be the ideal outcome, the approximations do appear to provide some meaningful information about the relative number of twins among different projection matrices. This is promising evidence that it may be feasible to compare node co-occurrence projection matrices with respect to their sets of twins without enumerating them, however we leave further exploration of this possibility for future work.

\subsection{Hypergraph Distances Among Sets of Twins}
\begin{figure*}[!ht]
    \centering
    \includegraphics[scale=0.4]{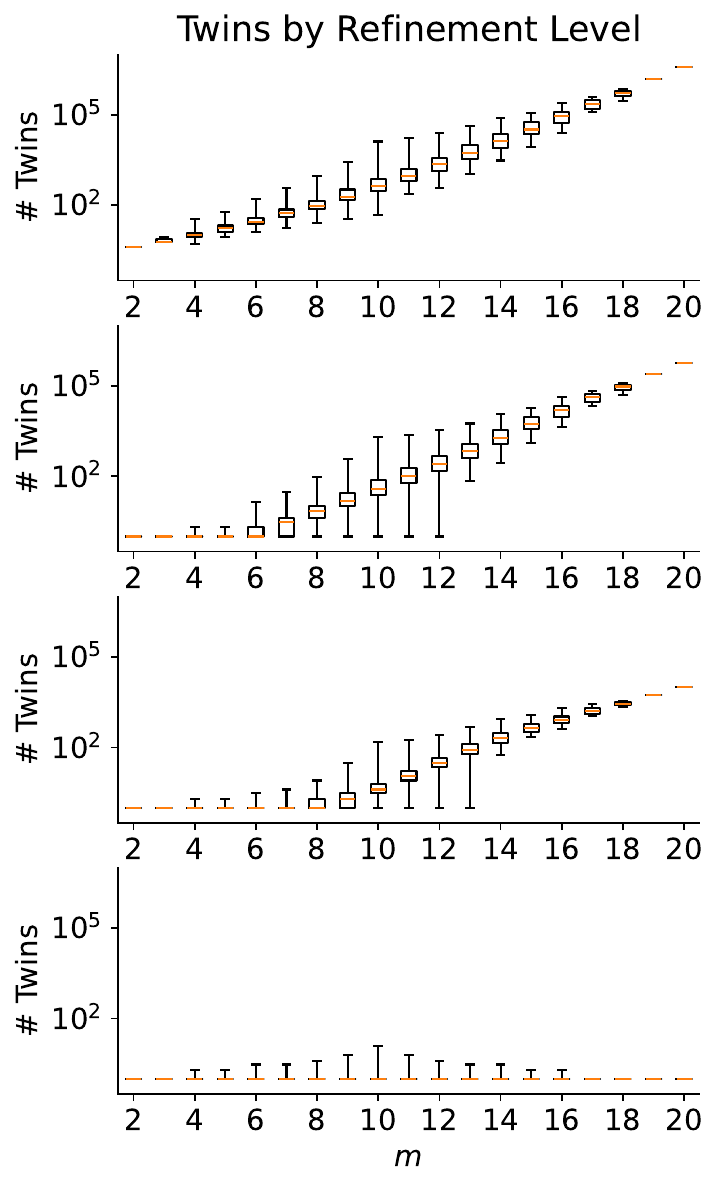}%
    \includegraphics[scale=0.4]{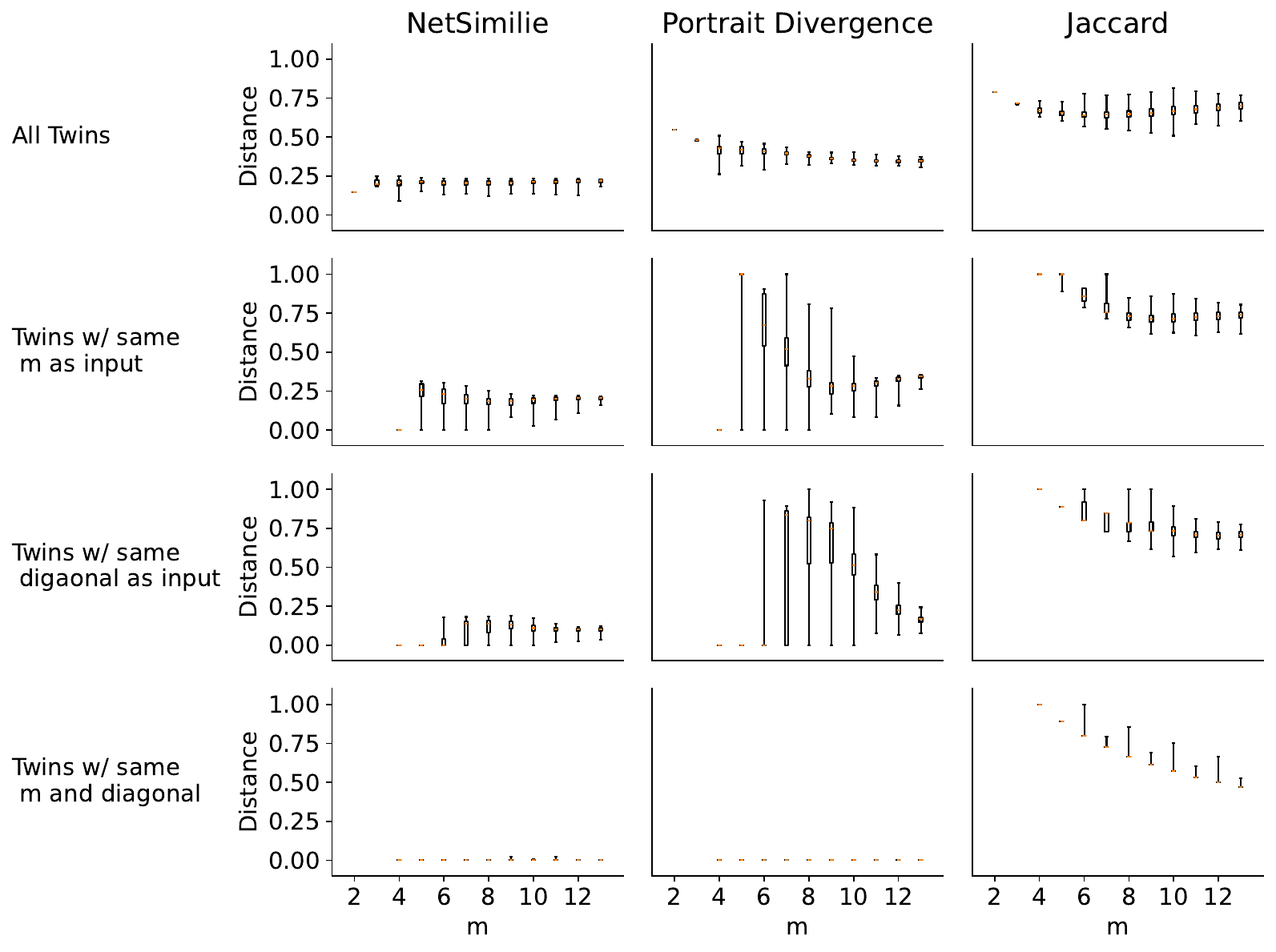}
    \caption{Left: Distribution of non-uniform twins per unique node co-occurrence projection at various levels of refinement. Each row shows a different level of refinement of the twin set, from all twins (top row) to only the twins with the same number of hyperedges and node hyperdegrees as the input (bottom row). Right: Distribution of the mean distance among twin hypergraphs for all unique up to isomorphism projections from $G_3(6,m)$.}
    \label{fig:distances}
\end{figure*}

We have so far characterized the set of twin hypergraphs associated to a node co-occurrence projection based primarily on the size of the set and the range of hyperedge size distributions. A natural next step is to evaluate the diversity of the hypergraphs within sets of twins. In this section, we use measures of pairwise hypergraph dissimilarity to analyze the diversity of sets of twins and connect twin set diversity back to properties of the node co-occurrence projection matrices.

We use three measures of hypergraph dissimilarity. First, to get a broad picture for the commonalities among the set of twin hypergraphs, we measure the Jaccard Index between their hyperedge sets. Given two sets $A$ and $B$, the Jaccard Index is measured as the difference between 1 and the intersection over the union of the two sets: $$1 - \frac{|A\cap B|}{|A \cup B|}\textrm{.}$$ A Jaccard Index of 1 indicates that there is no intersection between the two edge sets (\emph{i.e.}, they are disjoint), while a value of 0 indicates that the two sets are congruent. Since we have defined sets of twins such that the pairs of hypergraphs are distinct from one another, congruent sets are not possible in our setting, thus we know that the Jaccard Index between two twins will always be greater than 0. A value of 1 is possible, especially when comparing hypergraphs with different hyperedge size distributions, where there may be very little overlap between hyperedge sets.

We also compare two hypergraph dissimilarity measures recently adapted from pairwise network approaches by Agostinelli and colleagues \cite{agostinelli_higher-order_2025}. The first is an extension of the NetSimilie network comparison method \cite{berlingerio_network_2013}. This method embeds a hypergraph by first computing a feature vector for every node, then concatenating statistics of the node feature vectors to get an embedding vector for the hypergraph. Two hypergraphs are then compared by computing the Canberra distance between their feature vectors \cite{agostinelli_higher-order_2025}. The second measure is an extension of the Portrait Divergence for networks \cite{bagrow_information-theoretic_2019} to hypergraphs and is based on a 4-dimensional tensor that counts the number of hyperedges of size $m$ having $k$ hyperedges of size $n$ at distance $L$. This tensor is normalized into a probability distribution before the distance between two normalized portraits is computed using the Jensen-Shannon divergence \cite{agostinelli_higher-order_2025}.

There is a direct connection between our work and that of Agostinelli et al., as they use hypergraphs with the same pairwise projection to show why higher-order methods for hypergraph dissimilarity are needed beyond existing pairwise network approaches \cite{agostinelli_higher-order_2025}.\footnote{An important distinction is that their pairwise projection operator does not include the diagonal entries and is unweighted, whereas we consider weighted node co-occurrence matrices with diagonal entries.} They achieve this by designing a null model that begins from a reference hypergraph, then projects a random fraction of the hyperedges to pairwise interactions. They also formulate a second version of this null model from the opposite direction, beginning from a pairwise projection and randomly promoting a fraction of cliques of various sizes following the reference hyperedge size distribution into hyperedges. Both of these are potential ways of randomly generating hypergraphs from the unrefined set of twins. They apply this procedure to a real hypergraph, showing that as the fraction of random projections or promotions increases, so does the distance between the original and randomized hypergraphs, even while their pairwise projections remain equal. In contrast to these two complementary randomization procedures, our algorithm enumerates all of these hypergraphs into one set of twins, and in this section we evaluate how dissimilarity methods distinguish hypergraphs within the set of twins.\footnote{We note in passing that this approach is conceptually reminiscent of the within-ensemble graph distance proposed by Hartle and colleagues \cite{hartle_network_2020}.}

We again analyze the sets of twins based on our exhaustive search of $G_3(6,m)$. For each unique projection, we compute the distance between each pair of hypergraphs in its set of twins using the three measures described above. This gives us a pairwise distance distribution for every projection, which we then further divide based on refinements of the twin set. Due to the growth in the number of pairs of twin hypergraphs, as well as the computational complexity of the distance measures as implemented, we computed pairwise distances for sets of twins up to $m=13$.

We show results in Figure~\ref{fig:distances}. The left-hand plots shows the distribution of twin set sizes at various refinement levels. The unrefined set (top row) is the largest on average, followed by the set where we enforce the number of hyperedges to be exactly $m$, then the set of twins with matching hyperdegrees on the diagonal, and finally the set of twins with both of the previous two properties (bottom), where no projection has more than 100 twins remaining after refinement.

The right-hand box plots show the distribution of the average pairwise distance per projection at each refinement level. The hypergraph Netsimilie method stands out as the most consistent and intuitive of the three measures in the sense that it is not particularly sensitive to the number of hyperedges or the refinement level, and with the maximum amount of refinement, the average pairwise distance is 0. The Hypergraph Portrait Divergence shows similar results to NetSimilie for the unrefined and fully refined twin sets, but shows substantial variability for the two intermediate refinements. This is consistent with the finding in the original work that the Portrait Divergence is very sensitive to the hyperedge size distribution \cite{agostinelli_higher-order_2025}. Finally, the Jaccard distance follows a similar pattern across all refinements. It is consistently larger in magnitude than the other two measures, with a decrease in average distance as $m$ increases. In contrast to Hypergraph NetSimilie and Portrait Divergence, which both show dissimilarity values at or near 0 at the highest level of refinement, the Jaccard distributions remain relatively large and with some extreme values, reflecting the fact that the sets of hyperedges remain dissimilar even among hypergraphs that are ``similar'' according to their statistical and topological properties.

\subsection{Sampled k-uniform Search}
Beyond $G_3(6,m)$, it is generally not feasible to enumerate all hypergraphs or node co-occurrence projections as we did in the above analysis. Instead, in this section we run our algorithm on randomly sampled node co-occurrence projections. To guarantee that every projection matrix we sample is hypergraphical in the same way, we sample random projection matrices by first sampling a random hypergraph from $G_k(n,m)$ as defined in the beginning of this section, then computing its node co-occurrence projection.

This sampling procedure has two important implications. First, as stated above, it trivially guarantees that every sampled projection corresponds to at least one $k$-uniform hypergraph. Second, it implies that our results are likely to be overcounts, since we may sample the same projection multiple times either by sampling the same hypergraph repeatedly or by sampling different hypergraphs that have the same projection. This is especially likely when the binomial coefficient governing the number of possible hypergraphs is small relative to the number of samples. Similarly, our counts are also potentially inflated by isomorphism between projections, since two projections that are isomorphic will have the same number of twins. While overcounting is not ideal, the trade-off is that we can easily do the sampling and computations in parallel over a wider range of parameters than if we need to pairwise compare every sampled projection to all previous projections to ensure uniqueness.

In the remainder of this subsection we focus on the relationship between the properties of the random projections and the size of the set of $k$-uniform non-isomorphic twins. As discussed in Section~\ref{sec:methods}, to avoid extraneous computation and allow us to cover a wider range of parameters, we limit the factor graph to contain only $k$-cliques. However, the results would be the same if we computed all non-uniform twins, then post-processed the results to only the $k$-uniform twins.

\begin{figure}[!ht]
    \centering
    \includegraphics[scale=0.225]{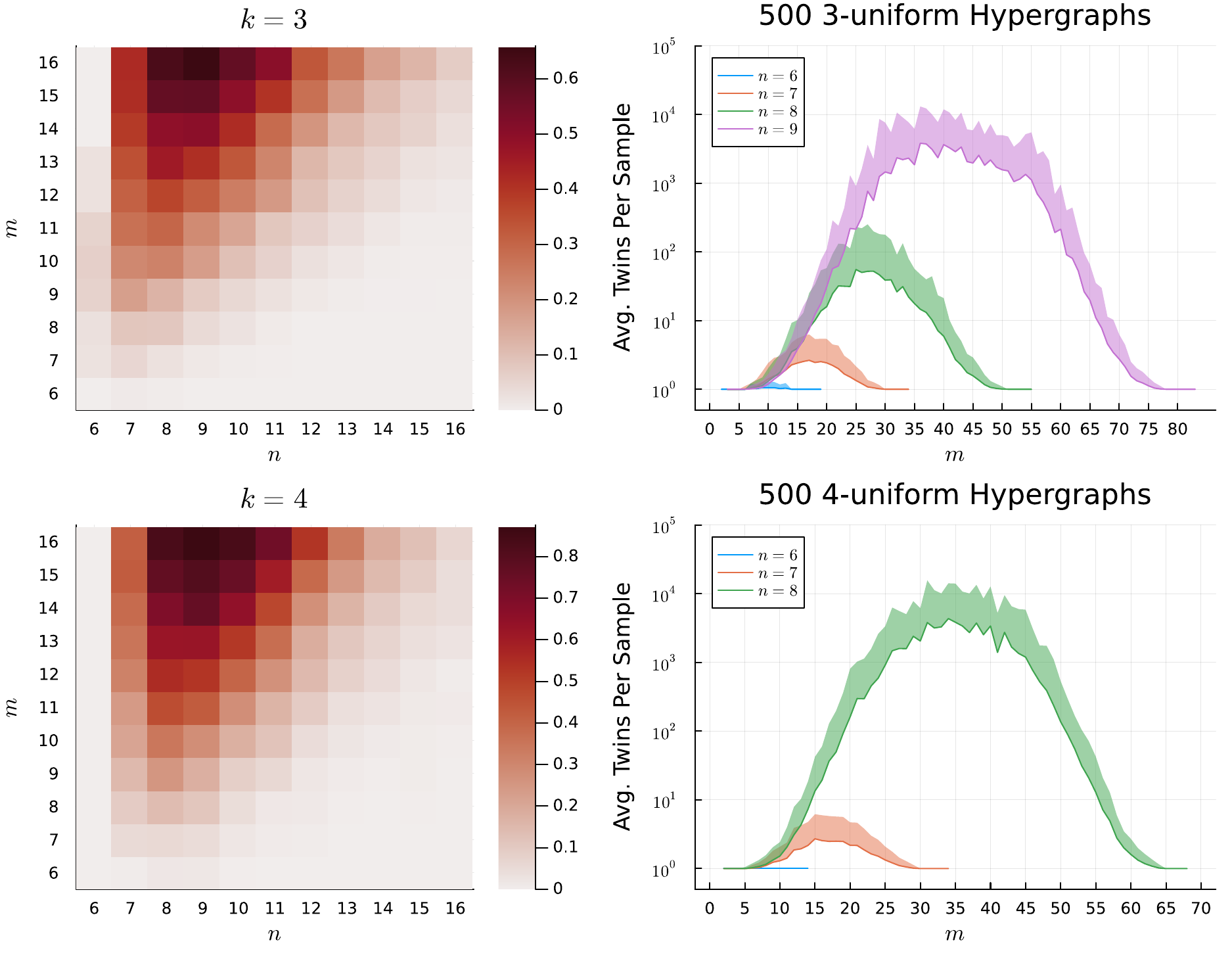}
    \caption{
    Left Column: Heatmaps showing proportion of 1000 projections sampled from $G_k(n,m)$ for $k \in \{3,4\}$ that have at least 2 $k$-uniform isomorphism classes in their set of non-isomorphic twins for varying values of $n$ (horizontal) and $m$ (vertical).
    Right: Average number of $k$-uniform isomorphism classes per sampled projection (500 samples) for varying values of $n$ (lines) up to maximum values of $m$ (horizontal axis) for simple hypergraphs.
    }
    \label{fig:3-uniform}
\end{figure}

In the left column of Figure~\ref{fig:3-uniform} we show results from varying the number of nodes $n$ and hyperedges $m$ from $6,\dots,16$ (similar to Salova and D'Souza \cite{salova_analyzing_2021}) for $k \in \{3,4\}$. Each cell in the heatmap corresponds to the proportion of 1000 sampled projections whose set of non-isomorphic twins has at least two $k$-uniform isomorphism classes. For both values of $k$ this proportion is largest around $n=9, m=16$ where more than 60\% of sampled projections correspond to at least two isomorphism classes of $k$-uniform hypergraph.

In the right column of Figure~\ref{fig:3-uniform}, we expand the columns of the heatmaps for $n\in 7,8,9$ for $k=3$ and $n\in 7,8$ for $k=4$. The plots show the average and standard deviation of the number of isomorphism classes per sampled projection across the entire hyperedge space for simple hypergraphs for each $k,n$ pair, \textit{i.e.}, when $\binom{\binom{n}{k}}{m} > 0$.

In general, as the number of nodes increases, so does the maximum number of twins per hypergraph. While the shape of the curves is similar, the order of magnitude of the average number of isomorphism classes increases substantially with $k$ and $n$, reaching averages above $10^3$ for $k=3,n=9$ and $k=4, n=8$. Further, the symmetry of the curves over the full range of $m$ indicates that increasing the density of the hypergraph increases the average size of the $k$-uniform set of non-isomorphic twins so long as it increases the number of possible hypergraphs, that is, while the binomial coefficient is increasing. This suggests that, similar to binomial coefficients in general, for a number of hyperedges $m$ the set of complements with $\binom{n}{k} - m$ hyperedges is likely to have similar properties. We leave further analysis of this dual relationship for future work.

\section{Conclusion}
In this paper we have defined and analyzed sets of twins associated to node co-occurrence projection matrices and presented an enumeration algorithm for their computation. We have related some properties of node co-occurrence projections with the size of the set of twins, and shown that a single projection may correspond to a wide variety of hypergraph structures, including hypergraphs that are not isomorphic with one another and those with different hyperedge size or hyperdegree distributions. These results, while limited in important ways, represent a substantial step towards a deeper, more systematic understanding of the information loss incurred by the use of node co-occurrence matrices for hypergraph analysis.

Many avenues for future research remain open. First, our results were limited to small and relatively simple hypergraph structures, with most results computed for hypergraphs on $n=6$ nodes up to a maximum of $n=9$, and the largest number of hyperedges $m=20$. Relatedly, all of the node co-occurrence projection matrices we studied were sampled based on a $k$-uniform hypergraph model with a homogeneous node hyperdegree distribution. This leaves open the possibility of studying larger hypergraph structures as well as more complex random models that explicitly incorporate features like hyperdegree and hyperedge size heterogeneity, node clustering, hyperedge overlap, and modular or community structure. 

The important question of the impact of the non-uniqueness of node co-occurrence projections on real datasets also remains open. Progress in this direction would allow us to evaluate directly the impact of non-uniqueness on methods that operate on the node co-occurrence projection. Since real group interaction datasets tend to be much larger than those we analyzed here, improvement over the basic enumerative approach, either through faster algorithms or, perhaps more likely, sampling or statistical inference methods, will likely be necessary to make analysis of real datasets feasible in general.

It may also be fruitful to further explore how properties of node co-occurrence matrices can be used to directly predict properties of their sets of twins. Here we have taken some first steps by evaluating the size of the search tree; the clique approximation; the number of twins, twin isomorphism classes, and pairs of Gram Mates; the diversity of hyperedge size distributions; and pairwise distances within sets of twins. However, we did not identify a single measure that can be computed from a node co-occurrence matrix and predict the complexity of the set of twins, nor did we find a single comparative measure that can correctly rank matrices by their twin set size without enumerating it. Such a measure could be used in conjunction with a null model to immediately evaluate, at least heuristically, whether the node co-occurrence matrix of a given hypergraph may be more or less susceptible to problems associated with non-uniqueness. It is clear from our results that properties such as the density of cliques in the projection matrix, the worst case search tree size, and the clique and hyperedge size distributions are all important factors, but we did not find any one of these factors to be strictly predictive.

We have also only analyzed hypergraph distances among sets of twins using a small selection of potential measures, which could be expanded in the future \cite{vasilyeva_distances_2023, qin_explainable_2023}. It may also be promising to explore the possibility of using sets of twins to design a within-ensemble hypergraph distance, similar to the distance defined by Hartle et al. \cite{hartle_network_2020}.

Finally, future work may focus specifically on the relationship between node co-occurrence matrices, sets of twins, and dynamical outcomes. Previous work has shown that the existence of twins matters \cite{salova_analyzing_2021, ferraz_de_arruda_phase_2021}. The ability to enumerate these twin hypergraphs opens the door for systematic approaches to understanding the impact of non-uniqueness on dynamics at both lower and higher orders.

\subsection*{Code \& Data Availability}
A {\CC} implementation of the TwinSearch procedure, as well as code for the analyses presented in this paper, will be made available on GitHub \cite{larock_TwinSearch_2023}.

\subsection*{Acknowledgments}
The authors acknowledge support from the EPSRC Grant  EP/V03474X/1. TL acknowledges Karel Devriendt, Brennan Klein, Maxime Lucas, Jelle Oostveen, Nicola Pedreschi, Rohit Sahasrabuddhe and Jason P. Smith for helpful conversations about the ideas in this paper. TL also acknowledges open source code made available by the developers of many projects including NumPy \cite{harris2020array}, SciPy \cite{2020SciPy-NMeth}, NetworkX \cite{SciPyProceedings_11}, MatPlotLib \cite{Hunter_2007}, and compleX Group Interactions (XGI) \cite{Landry_XGI_A_Python_2023}. TL acknowledges support from a grant from the Fund for Energy Research with Corporate Partners administered by the Andlinger Center for Energy and the Environment at Princeton University as well as the School of Engineering and Applied Science (SEAS) Seed Grant at Princeton University.

\printbibliography

\appendix

\section{Rank Correlation Coefficients}
\label{app:kt-stats}
In Table~\ref{app:tab:kt-stats}, we show for each value of $m$ the number of unique up to isomorphism node co-occurrence matrices, as well as the numerical values of Kendall's $\tau$ computed for the clique and worst case tree size approximations and the associated $p$-values corresponding to the data shown in the rightmost plot of Figure~\ref{fig:clique-approximation}.

\begin{table}[!ht]
    \centering
    \resizebox{\columnwidth}{!}{%
    \begin{tabular}{cccccc}
    \toprule
     & & \multicolumn{2}{c|}{Clique Approximation} & \multicolumn{2}{|c}{Worst Case Tree Size} \\
    \midrule
    $m$ & matrices & $\tau$ & $p$-value & $\tau$ & $p$-value \\
    \midrule
    3 & 3 & 1.000000 & 0.157299 & 1.000000 & 0.157299 \\
    4 & 15 & 0.878833 & 0.000012 & 0.717409 & 0.000511 \\
    5 & 37 & 0.600001 & 0.000000 & 0.535934 & 0.000005 \\
    6 & 88 & 0.533118 & 0.000000 & 0.494202 & 0.000000 \\
    7 & 155 & 0.427996 & 0.000000 & 0.383663 & 0.000000 \\
    8 & 243 & 0.403728 & 0.000000 & 0.331698 & 0.000000 \\
    9 & 303 & 0.432917 & 0.000000 & 0.312208 & 0.000000 \\
    10 & 342 & 0.461773 & 0.000000 & 0.322247 & 0.000000 \\
    11 & 304 & 0.535480 & 0.000000 & 0.335398 & 0.000000 \\
    12 & 245 & 0.605213 & 0.000000 & 0.347856 & 0.000000 \\
    13 & 159 & 0.663137 & 0.000000 & 0.363363 & 0.000000 \\
    14 & 94 & 0.696488 & 0.000000 & 0.366120 & 0.000000 \\
    15 & 43 & 0.740864 & 0.000000 & 0.371827 & 0.000761 \\
    16 & 21 & 0.647619 & 0.000011 & 0.433594 & 0.008701 \\
    17 & 7 & 0.809524 & 0.010714 & 0.514344 & 0.116852 \\
    18 & 3 & 1.000000 & 0.333333 & 0.000000 & 1.000000 \\
    \bottomrule
    \end{tabular}
    }
    \caption{Kendall's Tau rank correlation coefficients and p-values for points from Figure~\ref{fig:clique-approximation}.}
    \label{app:tab:kt-stats}
\end{table}

\end{document}